\documentclass{article}
\usepackage{graphicx}  
\usepackage{amsmath}   
\usepackage[compress]{cite}
\usepackage{amssymb}   
\usepackage{bm} 
\usepackage{dcolumn}
\usepackage{color}
\usepackage{mathrsfs}
\usepackage{amsfonts}
\usepackage{varioref}
\usepackage{textcomp}
\RequirePackage[colorlinks,citecolor=blue,urlcolor=magenta,linkcolor=blue]{hyperref}
\allowdisplaybreaks
\def\LL{Lanczos-Lovelock }

\def\gr{general relativity}

\newcommand{\ntabla}{\bar{\nabla}}
\labelformat{section}{Section #1} 
\labelformat{subsection}{Section #1} 
\labelformat{subsubsection}{Section #1}
\labelformat{subsubsubsection}{Section #1}
\labelformat{equation}{Eq.~(#1)} 
\labelformat{figure}{Fig.~#1} 
\labelformat{subfigure}{Fig.~\thefigure#1} 
\labelformat{table}{Tab.~#1} 
\labelformat{appendix}{Appendix #1}
\title{Noether Current, Black Hole Entropy and Spacetime Torsion}
\author{Sumanta Chakraborty \footnote{Email: sumantac.physics@gmail.com}~$^{1}$
and
Ramit Dey \footnote{Email: ramitdey@gmail.com}~$^{1,2}$
\\
$^{1}$\small{Department of Theoretical Physics, Indian Association for the Cultivation of Science, Kolkata 700032, India}
\\
$^{2}$\small{Department of Physics and Astronomy, The University of Western Ontario, London ON N6A 3K7, Canada}}
\begin{document}
  
\maketitle
\begin{abstract}
We show that the presence of spacetime torsion, unlike any other non-trivial modifications of the Einstein gravity, does \emph{not} affect black hole entropy. The theory being diffeomorphism invariant leads to a Noether current and hence to a Noether charge, which can be associated to the heat content of the spacetime. Furthermore, the evolution of the spacetime inheriting torsion can be encoded in the difference between suitably defined surface and bulk degrees of freedom. For static spacetimes the surface and bulk degrees of freedom coincides, leading to holographic equipartition. In doing so one can see that the surface degrees of freedom originate from horizon area and it is clear that spacetime torsion never contributes to the surface degrees of freedom and hence neither to the black hole entropy. We further show that the gravitational Hamiltonian in presence of torsion does not inherit any torsion dependence in the boundary term and hence the first law originating from the variation of the Hamiltonian, relates entropy to area. This reconfirms our claim that torsion does not modify the black hole entropy. 
\end{abstract}
\section{Introduction and Motivation}

Black holes have a characteristic temperature and entropy associated with them and this leads to the formulation of the laws of black hole mechanics \cite{Bardeen1973}. From the standpoint of Classical dynamics, this behaviour is a mystery, as there cannot exist any classical degrees of freedom to account for such thermodynamic properties, specially the temperature. Studying quantized matter fields in curved spacetime, it is possible to show that the black holes behave as a true thermodynamical system, reasoning the existence of the black hole temperature and entropy \cite{Bekenstein:1973ur,Hawking:1974sw,Gibbons:1976ue,Unruh:1976db}. This is essentially due to the fact that the vacuum states of quantum fields are not invariant under general co-ordinate transformation. While the quantum degrees of freedom of the matter fields can nicely account for the temperature associated with black holes, the notion of black hole entropy remained a mystery till date. 

The most likely explanation for the origin of black hole entropy is the microscopic quantum mechanical degrees of freedom hidden within the black hole horizon \cite{THOOFT1985727,Susskind:1993ws,Sen:1995in,Horowitz:1996qd,Strominger:1996sh,Rovelli:1996dv,Bombelli:1986rw,Jacobson:2005kr,Jacobson:2003wv}. The well-known Bekenstein-Hawking entropy associated with black holes in the context of general relativity takes the following form
\begin{align}
S_{\rm GR}=\frac{A_{\rm Horiozn}}{4\hbar G}.
\end{align}
As evident, the entropy depends on both the Planck's constant $\hbar$ and the Newton's constant $G$, pointing towards the tantalizing evidence that black hole thermodynamics unites quantum mechanics and gravity. This makes the computation of the correct black hole entropy in any quantum theory of gravity an important part of its foundation. Even though there are plethora of models, starting from which the above entropy-area relation can be derived, all these models are handicapped in one way or another. For example, the string theory calculation works only for supersymmetric or near-extremal black holes \cite{Strominger:1996sh,Callan:1995hn,Duff:1994an,Larsen:1995ss,Carlip:1995cd,Horowitz:1996qd}, while loop quantum gravity has to assume some choice for the ill-understood immirzi parameter \cite{Rovelli:1996dv,Ashtekar:1997yu,Date:2008rb,Frodden:2012dq,Pranzetti:2014tla,Neiman:2013ap}. Moreover, it was not at all clear if the above black hole entropy would remain the same by performing some non trivial modifications to  general relativity, such as including higher derivative terms in the action or dropping the assumption about a symmetric connection by including torsion as an additional degree of freedom.  

The modifications brought about in the entropy-area relation for black holes (or, for that matter generic null surfaces) is very important from the perspective of the gravity-thermodynamics connection. Recently, there have been several illuminating results shedding light on black hole thermodynamics. This started from the demonstration that one can arrive at Einstein's equations starting from Clausius relation \cite{Jacobson:1995ab}, which was supplemented by the demonstration that Einstein's equations can also be casted in a thermodynamic language \cite{Padmanabhan:2002sha}. Subsequently, both these methods were generalized to higher curvature theories and for generic null surfaces with modified expressions for entropy \cite{Padmanabhan:2003gd,Padmanabhan:2009vy,Kothawala:2009kc,Chakraborty:2015aja,Chakraborty:2015wma,Dey:2016zka,Sarkar:2013swa,Eling:2006aw}. In all these contexts the fact that the entropy-area relation is modified, played a crucial role in order to write down the respective expression associated with gravitational dynamics in a thermodynamic language. 

It is obvious that the invariance of the gravitational Lagrangian under diffeomorphism is a symmetry associated with the Lagrangian and hence there must be an associated Noether current. Given the Noether current it is straightforward to compute the Noether charge by integrating over a three-dimensional hypersurface. In the context of general relativity it turns out that the associated Noether charge can be exactly related to the black hole entropy and one recovers the entropy-area relation \cite{Wald:1993nt,Iyer:1994ys}. This prescription has been applied later on to various higher curvature theories and the associated entropy turns out to be different from area \cite{Jacobson:1993xs,Iyer:1994ys,Sheykhi:2007gi,Wu:2008ir,Kothawala:2007em,Chakraborty:2010wn,
Paranjape:2006ca,Kothawala:2009kc,Chakraborty:2014rga}. This has resulted in an understanding that whenever the Einstein-Hilbert Lagrangian is non-trivially modified, the entropy-area relation will also be affected. In a similar spirit the derivation of the first law for black hole spacetime also involves variation of the Noether current and associated symplectic structure of the spacetime. From the first law as well, one can immediately identify the expression for black hole entropy and this results into modifications of the entropy-area relation for theories beyond \gr\ \cite{Iyer:1994ys,Liberati:2017vse} (see also \cite{DePaoli:2018erh}). So far the above assertion has hold good for addition of higher derivative (or, higher curvature) terms in the action or modifications due to some non-minimal coupling of gravity with matter fields, leading to a departure from the entropy-area relation but not much has been said when one includes torsion in the theory and spacetime is no more Riemannian (however see \cite{Prabhu:2015vua,Barnich:2016rwk,Blagojevic:2006jk,Dey:2017fld}). Hence the above routes may help in understanding whether the entropy-area relation has been modified or not in the context of Riemann-Cartan spacetime. 

The interpretation of the Noether charge of any gravitational theory as the black hole entropy has further lead to some intriguing and deep understanding of the connection between gravitational dynamics and horizon thermodynamics. Recently in \cite{Padmanabhan:2013nxa,Chakraborty:2015hna,Chakraborty:2015wma} several additional results strengthening the above connection have been established within the framework of general relativity and Lanczos-Lovelock gravity as well. In particular, it was demonstrated in all these theories that the Noether charge associated with the diffeomorphism vector field generating time evolution is equal to the heat content of the boundary surface. In a similar spirit, it appears that evolution of the spacetime can be thought of as due to the difference between suitably defined surface and bulk degrees of freedom. As these two degrees of freedom become equal, the time evolution vector field becomes a Killing vector field and hence the spacetime ceases to be dynamical. In these scenarios as the theory of gravity is modified, the surface degrees of freedom responsible for black hole entropy also differs from being area. Thus the above derivation may also provide a handle on the modifications of the entropy-area relation for a diffeomorphism invariant theory other than \gr. In this work we will consider all these routes to asses the entropy-area relation in presence of spacetime torsion. 

The presence of spacetime torsion is unavoidable as quantum nature of the matter fields are taken into account. Most of the quantum fields enjoy their own intrinsic spin angular momentum. Therefore in the semi-classical framework, where one would like to extend general relativity in accordance with the underlying microphysics, inclusion of the spin angular momentum of matter is necessary and unavoidable. In the same way as the matter energy-momentum couples to the metric, `spin' couples to a geometrical quantity of spacetime related to rotational degrees of freedom and corresponds to non-trivial choices of the spacetime torsion \cite{Hehl:1976kj,Trautman:2006fp,Nester:2012ib,Chen:2015vya,Nester:2016rsy}. This results into a dramatic departure from the Riemannian description of the spacetime. Therefore in a complete description of gravity there is no escape from spacetime torsion, which is intrinsically geometrical and couples to spin angular momentum of matter fields \cite{Hehl:1976kj,Trautman:2006fp}. Further, the presence of spacetime torsion itself modifies the gravitational Lagrangian and hence the field equations in a \emph{non-trivial} manner. Even though some geometrical \cite{TRAUTMAN:1973aa,PhysRevD.10.1066} and thermodynamical \cite{Dey:2017fld} consequences of the presence of spacetime torsion has been discussed at some length, unfortunately there have been very little discussions on what happens to the black hole entropy or, more precisely to the entropy-area relation as torsion is brought into the picture \cite{Shapiro:2001rz,Hehl:1976kj} (as an aside, the effect of spacetime torsion on cosmology has been discussed in \cite{Shie:2008ms,Poplawski:2010kb,Banerjee:2018yyi}).  

In this work we firstly provide a brief review of the connection between Noether current and black hole entropy in \ref{NC_Review}. Subsequently we have provided a derivation of the Noether current for Einstein-Cartan theory, i.e., in a theory of gravity involving spacetime torsion as an additional degree of freedom in \ref{NC_Torsion}. Following the derivation of Noether current, in \ref{NC_Holography} we have explicitly demonstrated how the surface and bulk degrees of freedom splits. This further bring forth the fact that the presence of torsion does not effect the surface degrees of freedom at all, thus leading to the result that torsion would not effect the horizon entropy. It is further backed up by the derivation of the first law in the context of Einstein-Cartan spacetime using relevant Hamiltonian in \ref{NC_Hamiltonian}. We finally conclude with a discussion on the results in \ref{NC_Conclusion}.
\section{Noether Current to Black Hole Entropy: A Review of Earlier Results}
\label{NC_Review}

The correspondence between gravitational dynamics and spacetime thermodynamics transcends \gr\ and holds good for a variety of alternative gravitational theories, including \LL models of gravity. In both \gr\ and \LL models of gravity, the associated Noether current provides a bag full of thermodynamic relations, including an estimation of the black hole entropy (or, for that matter entropy associated with any null surface) in these theories. Therefore given any gravitational theory, it is possible to arrive at an expression for the Noether current associated with it and hence one may infer the associated black hole entropy. In this work we will employ an identical strategy to understand the black hole entropy as well as other thermodynamic identities associated with it in the presence of spacetime torsion. Since presence of spacetime torsion modifies the spacetime structure drastically it will be worthwhile to explore the earlier results in the context of \gr\ and \LL gravity to set up the background, with which we can compare and contrast our subsequent results. 

There can be two ways to arrive at the Noether current associated with \gr\ and \LL gravity, one of them follows from purely geometric point of view and is independent of the field equations, while the other one uses the gravitational field equations explicitly. Even though premises of both the Noether currents are different, due to very interesting cancellations they turn out to be identical. In the off-shell approach one considers the change in the gravitational action \emph{alone} due to diffeomorphism and hence determines the Noether current. On the other hand, the on-shell approach considers the variation of $(\textrm{gravity}+\textrm{matter})$ action, which leads to the Noether current when gravitational field equations are imposed. For matter action involving minimally coupled scalar or electromagnetic field both these currents match. In the context of \gr\ the Noether current associated with a diffeomorphism vector field $v^{\alpha}$ takes the following form, 
\begin{align}
J^{\mu}[v]
&=\frac{1}{16\pi G}\nabla _{\nu}\left\{\nabla ^{\mu}v^{\nu}-\nabla ^{\nu}v^{\mu}\right\}
\label{NC_Review_01}
\\
&=\frac{1}{8\pi G}R^{\mu}_{\nu}v^{\nu}+\frac{1}{16\pi G}g^{\alpha \beta}\pounds _{v}N^{\mu}_{\alpha \beta}
\label{NC_Review_02}
\end{align}
where $N^{\mu}_{\alpha \beta}\equiv -\Gamma ^{\mu}_{\alpha \beta}+(1/2)(\delta ^{\mu}_{\alpha}\Gamma ^{\rho}_{\rho \beta}+\delta ^{\mu}_{\beta}\Gamma ^{\rho}_{\rho \alpha})$ \cite{Parattu:2013gwa}. The above Noether current inherits intriguing thermodynamic relations, which becomes apparent when one introduces the following setup: Introduction of a time coordinate $t$ foliating the spacetime, such that we have a hypersurface orthogonal vector field $u_{\alpha}=-N\nabla _{\alpha}t$, where $N$ is the normalization factor. Then the time evolution vector field in this context, which would become Killing in static situations become $\xi^{\mu}=Nu^{\mu}$. Interestingly the Noether charge density associated with $\xi^{\mu}$ on a $t=\textrm{constant}$ hypersurface becomes,
\begin{align}
16\pi G~u_{\mu}J^{\mu}[\xi]=D_{\alpha}\left\{2Na^{\alpha}\right\}
\label{NC_Review_03}
\end{align}
where $a^{\alpha}$ is the acceleration associated with $u_{\mu}$ and $D_{\alpha}$ is the covariant derivative on the $t=\textrm{constant}$ hypersurface. Thus when integrated over a three-surface on the $t=\textrm{constant}$ hypersurface, the term on the right hand side will be converted to an integral over a two-surface such that,
\begin{align}
\int _{\mathcal{V}}d^{3}x~\sqrt{h}~u_{\mu}J^{\mu}[\xi]
=\epsilon \int _{\partial \mathcal{V}}d^{2}x~\left(\frac{Na}{2\pi}\right)\frac{\sqrt{\sigma}}{4G}
=\epsilon\int _{\partial \mathcal{V}} d^{2}x~T_{\rm loc}s
\label{NC_Review_04}
\end{align}
where $\epsilon$ is a numerical factor taking values $+1(\textrm{or,}-1)$ depending on whether the normal to the two-surface $\partial \mathcal{V}$ is outward (or, inward) pointing. Here the two terms on the right hand side has the following interpretations --- (a) one of them will be $\sqrt{\sigma}/4G$, the entropy density $s$ and (b) $Na/2\pi$, the Davies-Unruh redshifted temperature $T_{\rm loc}$ associated with the acceleration of the observer with four-velocity $u^{\mu}$ \cite{Davies:1974th,Unruh:1976db}. Thus the Noether charge identifies $\textrm{area}/4$ as the entropy associated with black holes (or, null surfaces) in \gr.

The above result can be used to determine an expression for spacetime evolution as well, which corresponds to $g^{\alpha \beta}\pounds _{v}N^{\mu}_{\alpha \beta}$ term. For static spacetimes, $\xi ^{\mu}$ will be the Killing vector field associated with temporal isometry and hence this Lie variation term will vanish. Using Einstein's equations in the form, $R_{\mu \nu}=8\pi G \{T_{\mu \nu}-(1/2)Tg_{\mu \nu}\}$, we obtain the following expression for the Lie variation term,
\begin{align}
\frac{1}{8\pi G}\int _{\mathcal{V}}d^{3}x~\sqrt{h}~u_{\mu}g^{\alpha \beta}\pounds _{\xi}N^{\mu}_{\alpha \beta}=\frac{1}{2}T_{\rm avg}\left\{N_{\rm sur}-N_{\rm bulk}\right\}
\label{NC_Review_05}
\\
N_{\rm sur}=\frac{\textrm{Area}}{G};\qquad N_{\rm bulk}=\frac{1}{(1/2)T_{\rm avg}}\int _{\mathcal{V}}d^{3}x~\sqrt{h}~\rho_{\rm Komar}
\label{NC_Review_06}
\end{align}
Here $T_{\rm avg}$ is the average temperature over the two-surface $\partial \mathcal{V}$ and $\rho_{\rm Komar}=\{T_{\mu \nu}-(1/2)Tg_{\mu \nu}\}u^{\mu}u^{\nu}$ is the Komar energy density. Thus the difference between suitably defined surface and bulk degrees of freedom is responsible for the Lie variation term to be non-zero and hence evolution of spacetime. If these two degrees of freedom coincide then the spacetime is static as the Lie variation vanishes and $\xi^{\mu}$ becomes a Killing vector field. 
\section{Noether Current in Presence of Spacetime Torsion}
\label{NC_Torsion}

In this section we will determine the Noether current associated with the diffeomorphism invariance of the gravitational action albeit in the presence of spacetime torsion. The first place where spacetime torsion plays a significant role is in the definition of covariant derivative. In general, there is no such restriction for the connection appearing in the covariant derivative to be symmetric and hence it can inherit anti-symmetric parts, which is known as torsion tensor $T^{\mu}_{~\alpha \beta}$. However the additional part of the connection modulo the Christoffel symbol is known as the contorsion tensor $K^{\mu}_{~\alpha \beta}$. The torsion and the contorsion tensors can be related by demanding that the covariant derivative of the metric tensor still vanishes even in presence of torsion. In this case the gravitational Lagrangian is again taken to be the Ricci scalar but obtained from the Riemann-Cartan manifold \cite{Arcos:2005ec}, which can be written as,
\begin{align}
L_{\rm EC}=\bar{R}&=R-\left(T_{\rho}T^{\rho}-K^{\alpha \mu \rho}K_{\mu \alpha \rho}\right)
+2\nabla _{\alpha}T^{\alpha}
\label{NC_Torsion_01}
\\
&=R+2\ntabla _{\alpha}T^{\alpha}+T_{\rho}T^{\rho}+K^{\alpha \rho \mu}K_{\rho \alpha \mu}
\label{NC_Torsion_02}
\end{align}
where the subscript `EC' stands for Einstein-Cartan theory. The quantity $T_{\rho}$ corresponds to the trace of the torsion tensor, which is defined as $T_{\rho}\equiv T^{\mu}_{~\rho \mu}=-K^{\mu}_{~\mu \rho}$, where $K^{\mu}_{~\alpha \beta}$ is the contorsion tensor. The above Lagrangian can also be neatly separated into a surface part and a bulk part, where the bulk part will inherit contributions from torsion, but not the surface part. Thus in this context as well the holographic relation between the bulk and boundary action of Einstein-Cartan theory does not hold (for another example, see \cite{Banerjee:2010yn}).

Note that the above Lagrangian as presented in \ref{NC_Torsion_01} has a total derivative term and hence the field equation with or without the $\nabla _{\alpha}T^{\alpha}$ term remains the same. This essentially amounts to an ambiguity in the expression for Noether current. Given the above action, one has to follow the usual procedure, i.e., vary the gravitational action under a diffeomorphism, which will eventually lead to the Noether current. Interestingly, unlike the scenarios described in \ref{NC_Review}, derivation of the Noether current from off-shell approach in the Einstein-Cartan theory is complicated. This is mainly due to the fact under diffeomorphism both the metric and the contorsion (or, torsion) will change. Even though the variation of the metric can be written in terms of derivatives of the diffeomorphism vector field, the variation of the contorsion tensor has no such simple expression. This prohibits one to determine the Noether current in Einstein-Cartan theory using the off-shell method. Note that for actions involving non-minimal coupling one must take into account the non-minimal terms as well, in which case it is known that there can be ambiguities between the off-shell and on-shell methods. As a curiosity, we would like to mention that even in the absence of non-minimal coupling, the on-shell and off-shell methods do \emph{not} give identical expressions for Noether current. Such a scenario corresponds to a particular class of Galileon Lagrangian. Given the above difficulty with the off-shell method we will consider the on-shell approach. In this case knowledge about the gravitational Lagrangian is not enough, one needs to know the matter sector as well. 

In the Einstein-Cartan theory, torsion is not dynamical and hence in absence of a source term the torsion tensor identically vanishes, resulting into normal general relativistic scenario. Thus it is necessary to introduce a source term for torsion, which may have its origin from spin fluid, which reads, $-K^{\alpha}_{~\sigma \beta}\tau ^{\sigma~\beta}_{~\alpha}$. Here $\tau ^{\sigma~\beta}_{~\alpha}$ is the generator of the spin fluid. Similarly we assume an additional minimally coupled scalar field in the matter sector. Thus the complete action of $(\textrm{gravity}+\textrm{matter})$ takes the following form,
\begin{align}
\mathcal{A}_{\rm tot}&=\int d^{4}x \sqrt{-g}\left(L_{\rm EC}+L_{\rm matter}\right)=\int d^{4}x \sqrt{-g}\Big[\frac{\bar{R}}{16\pi G}-K^{\alpha}_{~\sigma \beta}\tau ^{\sigma~\beta}_{~\alpha}
-\frac{1}{2}\nabla _{\alpha}\Phi \nabla ^{\alpha}\Phi -V(\Phi)\Big]
\label{NC_Torsion_03}
\end{align}
where, $\tau ^{\mu}_{~\alpha \beta}$ acts as the source of the torsion tensor and $K^{\alpha}_{~\mu \nu}$ is the contorsion tensor as explained earlier. Also $\bar{R}$ denotes the Ricci scalar in presence of spacetime torsion. Variation of the above action with respect to metric, contorsion tensor and the scalar field yields,
\begin{align}
\delta \mathcal{A}_{\rm tot}&=\int d^{4}x\sqrt{-g}\Big[\frac{1}{16\pi G}E_{\mu \nu}\delta g^{\mu \nu}
+\frac{1}{16\pi G}\chi ^{\sigma~\beta}_{~\alpha}\delta K^{\alpha}_{~\sigma \beta}+E_{(\Phi)}\delta \Phi+\nabla _{\mu}\delta V^{\mu} \Big]
\label{NC_Torsion_04}
\end{align}
The above variation involves four terms, the first three are due to variations of the independent variables and the last one is the total derivative term. Given the above one can read off the gravitational field equations by setting the coefficient of the dynamical variable i.e., the metric to be vanishing. This yields $E_{\mu \nu}=0$ and takes the following form,
\begin{align}
\bar{G}_{\mu \nu}-\left(\bar{\nabla}_{\alpha}+T_{\alpha}\right)\left\{K^{\alpha}_{~\mu \nu}+\delta ^{\alpha}_{\mu}T_{\nu}-g_{\mu \nu}T^{\alpha} \right\}&=8\pi G\left\{\nabla _{\mu}\Phi \nabla _{\nu}\Phi +g_{\mu \nu}\left(-\frac{1}{2}\nabla _{\alpha}\Phi \nabla ^{\alpha}\Phi -V(\Phi) \right) \right\}
\nonumber
\\
&+8\pi G \left\{K^{\alpha}_{~\sigma \mu}\tau ^{\sigma}_{~\alpha \nu}+K^{\alpha}_{~\sigma \nu}\tau ^{\sigma}_{~\alpha \mu}
-g_{\mu \nu}\left(K^{\alpha}_{~\sigma \beta}\tau ^{\sigma ~ \beta}_{~\alpha}\right) \right\}
\label{NC_Torsion_05}
\end{align}
where, $\bar{G}_{\mu \nu}$ is the Einstein tensor in presence of spacetime torsion. This in turn can be written in terms of the Einstein tensor alone, by relating quantities in presence of torsion to those in its absence. The above exercise finally yields, 
\begin{align}
G_{\mu \nu}&-T_{\mu}T_{\nu}+K^{\alpha \beta}_{~~\mu}K_{\beta \alpha \nu}+\frac{1}{2}g_{\mu \nu}\left(T_{\alpha}T^{\alpha}
-K_{\alpha \beta \rho}K^{\beta \alpha \rho}\right)
\nonumber
\\
&=8\pi G\left\{\nabla _{\mu}\Phi \nabla _{\nu}\Phi +g_{\mu \nu}\left(-\frac{1}{2}\nabla _{\alpha}\Phi \nabla ^{\alpha}\Phi -V(\Phi) \right) \right\}+8\pi G \left\{K^{\alpha}_{~\sigma \mu}\tau ^{\sigma}_{~\alpha \nu}+K^{\alpha}_{~\sigma \nu}\tau ^{\sigma}_{~\alpha \mu}
-g_{\mu \nu}\left(K^{\alpha}_{~\sigma \beta}\tau ^{\sigma ~ \beta}_{~\alpha}\right) \right\}
\label{NC_Torsion_06}
\end{align}
In an identical manner the field equations for the torsional degrees of freedom can also be determined by equating the coefficient of $\delta K^{\mu}_{~\alpha \beta}$ to zero. This leads to the following equation satisfied by the torsional degrees of freedom: $\chi ^{\mu}_{~\alpha \beta}=0$. On expanding out in terms of torsion tensor and its trace, the above equation results into,
\begin{align}
S^{\mu}_{~\alpha \beta}\equiv T^{\mu}_{~\alpha \beta}+T_{\beta}\delta ^{\mu}_{\alpha}
-T_{\alpha}\delta ^{\mu}_{\beta}=16\pi G \tau ^{\mu}_{~\alpha \beta}
\label{NC_Torsion_07}
\end{align}
The field equation for the scalar field is the standard one and reads $E_{\Phi}=\square \Phi-(\partial V/\partial \Phi)=0$. Finally the boundary contribution has the following structure, 
\begin{align}
\delta V^{\mu}=\frac{1}{16\pi G}g^{\mu \nu}g^{\alpha \beta}\left(\nabla _{\alpha}\delta g_{\nu \beta}-\nabla _{\nu}\delta g_{\alpha \beta}\right)-\nabla ^{\mu}\Phi \delta \Phi
\label{NC_Torsion_08}
\end{align}
At this stage the field equations are assumed to hold true, so that the first three terms appearing in \ref{NC_Torsion_04} drops out. Further, if the above variation is taken to be due to diffeomorphism by a vector field $v^{\mu}$, such that $x^{\mu}\rightarrow x^{\mu}+v^{\mu}$, we obtain, 
\begin{align}
\delta _{v}\mathcal{A}_{\rm tot}=-\int d^{4}x~\sqrt{-g}\nabla _{\mu}\left\{\left(L_{\rm EC}+L_{\rm matter}\right)v^{\mu}\right\}
\label{NC_Torsion_09}
\end{align}
Thus when respective field equations are satisfied it follows that under diffeomorphism, variation of the total action essentially corresponds to covariant conservation of a four-current $J^{\mu}$, such that $\nabla _{\mu}J^{\mu}=0$. This is known as the on-shell Noether current and has the following expression,
\begin{align}
J^{\mu}[v]&=\delta _{v}V^{\mu}+\left\{\frac{\bar{R}}{16\pi G}-K^{\alpha}_{~\sigma \beta}\tau ^{\sigma~\beta}_{~\alpha}
-\frac{1}{2}\nabla _{\alpha}\Phi \nabla ^{\alpha}\Phi -V(\Phi)\right\}v^{\mu}
\label{NC_Torsion_10}
\end{align}
The Lie variation of the boundary term involves $\delta _{v}g_{\alpha \beta}$ and $\delta _{v}\Phi$ as evident from \ref{NC_Torsion_08}. These terms can be written as --- (a) $\delta _{v}\Phi=-v^{\alpha}\nabla _{\alpha}\Phi$ and (b) $\delta _{v}g_{\alpha \beta}=\nabla _{\alpha}v_{\beta}+\nabla _{\beta}v_{\alpha}$. Since Lie variation involves partial derivatives, they are not affected by the presence of torsion. Also note that in the above we have not taken into account the total derivative term present in the gravitational action and thus the Noether current presented above will have that ambiguity. Using the above expressions for Lie variation of metric and the scalar field we obtain, 
\begin{align}
J^{\mu}[v]&=\frac{1}{16\pi G}\nabla _{\alpha}\left(\nabla ^{\mu}v^{\alpha}
-\nabla ^{\alpha}v^{\mu} \right)
+v^{\mu}\left\{-K^{\alpha}_{~\sigma \beta}\tau ^{\sigma~\beta}_{~\alpha}
+\frac{1}{16\pi G}\left(-T_{\rho}T^{\rho}+K^{\alpha \mu \rho}K_{\mu \alpha \rho}\right) \right\}
\nonumber
\\
&+v^{\nu}\left\{\nabla ^{\mu}\Phi \nabla _{\nu}\Phi +\delta ^{\mu}_{\nu}\left(-\frac{1}{2}\nabla _{\alpha}\Phi \nabla ^{\alpha}\Phi -V(\Phi) \right) \right\}
-\frac{1}{8\pi G}\left(R^{\mu}_{\nu}-\frac{1}{2}\delta ^{\mu}_{\nu}R\right)v^{\nu}
\label{NC_Torsion_11}
\end{align}
Since the above Noether current is derived on-shell, we can use field equations to simplify it further. In particular, the last term in the above expression can be replaced by using the gravitational field equations, presented in \ref{NC_Torsion_06}. Thus we finally arrived at the following expression for the on-shell Noether current,
\begin{align}
J^{\mu}[v]&=\frac{1}{16\pi G}\nabla _{\alpha}\left(\nabla ^{\mu}v^{\alpha}
-\nabla ^{\alpha}v^{\mu} \right)
-v^{\nu}\left\{K^{\alpha}_{~\sigma \mu}\tau ^{\sigma}_{~\alpha \nu}+K^{\alpha}_{~\sigma \nu}\tau ^{\sigma}_{~\alpha \mu}\right\}
-\frac{1}{8\pi G}\Big\{T_{\mu}T_{\nu}-K^{\alpha \beta}_{~~\mu}K_{\beta \alpha \nu}\Big\}v^{\nu}
\label{NC_Torsion_12}
\end{align}
The above depicts the expression for the Noether current in presence of spacetime torsion. The only problematic feature of the above expression is associated with its mixed nature, i.e., the contorsion tensor and $\tau ^{\mu}_{~\alpha \beta}$ are both present. However given Eq.(13) one can either write down the Noether current explicitly in terms of the contorsion tensor or in terms of the torsion. In the first case use of the below identities 
\begin{align}
K^{\alpha \sigma \mu}\tau _{\sigma \alpha \nu}&=\frac{1}{16\pi G}\left\{K^{\alpha}_{~\sigma \mu}K^{\sigma}_{~\alpha \nu}+K^{\alpha}_{~\sigma \mu}K_{\alpha \nu}^{~~~\sigma}
-T_{\mu}T_{\nu}-T_{\alpha}K^{\alpha}_{~\nu \mu}\right\}
\label{NC_Torsion_13a}
\\
K^{\alpha \sigma \nu}\tau _{\sigma \alpha \mu}&=\frac{1}{16\pi G}\left\{K^{\alpha}_{~\sigma \nu}K^{\sigma}_{~\alpha \mu}+K^{\alpha}_{~\sigma \nu}K_{\alpha \mu}^{~~~\sigma}-T_{\nu}T_{\mu}-T_{\alpha}K^{\alpha}_{~\mu \nu}\right\}
\label{NC_Torsion_13b}
\end{align}
lead to the following expression for the on-shell Noether current as, 
\begin{align}
J^{\mu}[v]&=\frac{1}{16\pi G}\nabla _{\alpha}\left(\nabla ^{\mu}v^{\alpha}
-\nabla ^{\alpha}v^{\mu} \right)
-\frac{1}{16\pi G}v^{\nu}\Big\{K^{\alpha\sigma \mu}K_{\alpha \nu \sigma }+K^{\alpha \mu \sigma}K_{\alpha \sigma \nu}
-T_{\alpha}\left(K^{\alpha}_{~\nu \mu}+K^{\alpha}_{~\mu \nu}\right)\Big\}
\label{NC_Torsion_14}
\\
&=J^{\mu}_{\rm gr}[v]+J^{\mu}_{\rm tor}[v]
\label{NC_Torsion_N1}
\end{align}
It is also possible to write down the above expression for Noether current explicitly in terms of the torsion tensor by using the expansion of the contorsion tensor in terms of the torsion. This results into the following two identities,
\begin{align}
K^{\alpha \mu \sigma }K_{\alpha \sigma \nu}&+K^{\alpha \sigma \mu}K_{\alpha \nu \sigma}
=\frac{1}{2}T^{\mu \alpha \sigma}T_{\nu \alpha \sigma}+\frac{1}{2}T^{\alpha \mu  \sigma}T_{\nu \alpha \sigma}
+\frac{1}{2}T^{\mu \alpha \sigma}T_{\alpha \nu \sigma}
\label{NC_Torsion_15}
\\
K^{\alpha \mu}_{~~~\nu}&+K^{\alpha~\mu}_{~\nu}=T^{\mu \alpha}_{~~~\nu}+T_{\nu}^{~\alpha \mu}
\label{NC_Torsion_16}
\end{align}
Using these two identities, we obtain  the Noether current in terms of spacetime torsion as,
\begin{align}
J^{\mu}[v]&=\frac{1}{16\pi G}\nabla _{\alpha}\left(\nabla ^{\mu}v^{\alpha}-\nabla ^{\alpha}v^{\mu} \right)
\nonumber
\\
&-\frac{1}{16\pi G}v^{\nu}\Big\{\frac{1}{2}T^{\mu \alpha \sigma}T_{\nu \alpha \sigma}+\frac{1}{2}T^{\alpha \mu  \sigma}T_{\nu \alpha \sigma}
+\frac{1}{2}T^{\mu \alpha \sigma}T_{\alpha \nu \sigma}
-T_{\alpha}\left(T^{\mu \alpha}_{~~~\nu}+T_{\nu}^{~\alpha \mu}\right)\Big\}
\label{NC_Torsion_17}
\end{align}
Even though all these expressions have been presented for completeness, for our later purposes we will use the expression dependent on contorsion tensor as given by \ref{NC_Torsion_14}. This will help to simplify the algebra significantly. Note that when torsion tensor vanishes we get back \ref{NC_Review_01}, the Noether current for \gr. This acts as a consistency check of our formalism. Having described the Noether current in presence of spacetime torsion, we turn our attention to its thermodynamical significance in the subsequent sections. 
\section{Holographic Equipartition and Spacetime Evolution in Einstein-Cartan Theory}
\label{NC_Holography}

In this section we will show that even in the context of Einstein-Cartan theory it is possible to have a holographic equipartition by separating out a suitable bulk and surface degrees of freedom. Moreover the difference between the bulk and surface degrees of freedom can be held responsible for spacetime evolution just as in the case of general relativity \cite{Padmanabhan:2013nxa}. 

For this purpose, we start with the gravitational field equations in presence of spacetime torsion as elaborated in \ref{NC_Torsion_06}. Given the Einstein tensor, the Ricci scalar can be readily obtained by taking the trace of \ref{NC_Torsion_06}, which ultimately yields,
\begin{align}
R=8\pi G \left\{\nabla _{\alpha}\Phi \nabla ^{\alpha}\Phi +4V(\Phi) \right\}
+16\pi G K^{\alpha \sigma \beta}\tau _{\sigma \alpha \beta}
+\left(T_{\alpha}T^{\alpha}-K_{\alpha \beta \rho}K^{\beta \alpha \rho}\right)
\label{HE_Torsion_01}
\end{align}
Therefore, the Ricci tensor can also be determined by substituting the expression for Ricci scalar from \ref{HE_Torsion_01} to that of the Einstein tensor, resulting into,
\begin{align}
R_{\mu \nu}&=8\pi G \left\{\nabla _{\mu}\Phi \nabla _{\nu}\Phi +g_{\mu \nu}V(\Phi)\right\}
+8\pi G \left\{K^{\alpha \sigma}_{~~~\mu}\tau _{\sigma \alpha \nu}+K^{\alpha \sigma}_{~~~\nu}\tau _{\sigma \alpha \mu}\right\}
+T_{\mu}T_{\nu}-K^{\alpha \beta}_{~~~\mu}K_{\beta \alpha \nu}
\label{HE_Torsion_02}
\end{align}
The above equation is similar to the Einstein's equations provided one interprets the right hand side to be the matter energy momentum tensor. Since the origin of the field $\Phi$ and the torsion (or, contorsion) field $T^{\mu}_{~\alpha \beta}$ are very different, the right hand side consists of two independent contributions, one form the scalar field and the other from the contorsion field. Thus it is legitimate to define,
\begin{align}
\bar{T}_{\mu \nu}^{\rm (matter)}&=\left\{\nabla _{\mu}\Phi \nabla _{\nu}\Phi +g_{\mu \nu}V(\Phi)\right\}
\label{HE_Torsion_03a}
\\
\bar{T}_{\mu \nu}^{\rm (torsion)}&=\left\{K^{\alpha \sigma}_{~~~\mu}\tau _{\sigma \alpha \nu}+K^{\alpha \sigma}_{~~~\nu}\tau _{\sigma \alpha \mu}\right\}
+T_{\mu}T_{\nu}-K^{\alpha \beta}_{~~~\mu}K_{\beta \alpha \nu}
\label{HE_Torsion_03b}
\end{align}
This sets up the basic results associated with the evolution of the spacetime via the gravitational field equations. We will now use these relations to manipulate the expression for the Noether current to arrive at an equipartition relation for spacetimes inheriting a timelike Killing vector field. 

The Noether current, by definition, is conserved and hence one can also construct a conserved Noether charge associated with any vector field acting as the generator of diffeomorphism. However when this vector field relates to the time evolution of the spacetime it becomes of significant interest. To understand the time evolution let us foliate the spacetime by $t=\textrm{constant}$ hypersurface, with the normal being proportional to $\nabla _{\mu}t$. Given this setup it turns out that the vector field $\xi ^{\mu}=Nu^{\mu}$, where $u_{\mu}=-N\nabla _{\mu}t$ is the normalized vector field orthogonal to $t=\textrm{constant}$ hypersurface, is intimately connected with the time evolution of the spacetime \cite{Padmanabhan:2013nxa}. Note that in static spacetime $\xi ^{\mu}$ is the Killing vector field generating time translation. 

Surprisingly enough, it turns out that even in the presence of spacetime torsion it is possible to provide a simple algebraic relation for the Noether charge, which will have direct thermodynamic significance. To compute the total Noether charge within a three-volume on a $t=\textrm{constant}$ hypersurface, we start by projecting the Noether current $J^{\mu}[\xi]$ along $u^{\mu}$. This yields the Noether charge density and hence the Noether charge itself can be computed by integrating over a 3-volume on the $t=\textrm{constant}$ surface with $\sqrt{h}$ as the integration measure, where $h$ is the determinant of the induced metric on the $t=\textrm{constant}$ hypersurface. Thus we obtain, the Noether charge density to have the following form,
\begin{align}
16\pi G~u_{\mu}J^{\mu}[\xi]&=2NR_{\mu \nu}u^{\mu}u^{\nu}+u_{\mu}g^{\alpha \beta}\pounds _{\xi}N^{\mu}_{\alpha \beta}
-32\pi G~NK^{\alpha \sigma}_{~~~\mu}\tau _{\sigma \alpha \nu}u^{\mu}u^{\nu}
-N\left(-T_{\alpha}T^{\alpha}+K_{\alpha \beta \rho}K^{\beta \alpha \rho}\right)
\nonumber
\\
&-2N\left\{T_{\mu}T_{\nu}-K^{\alpha \beta}_{~~\mu}K_{\beta \alpha \nu}-\frac{1}{2}g_{\mu \nu}\left(T_{\alpha}T^{\alpha}
-K_{\alpha \beta \rho}K^{\beta \alpha \rho}\right)\right\}u^{\mu}u^{\nu}
\label{HE_Torsion_04}
\end{align}
Here we have used the following identity, $\nabla_{\alpha}(\nabla ^{\mu}v^{\alpha}-\nabla ^{\alpha}v^{\mu})=2R^{\mu}_{\nu}v^{\nu}+g^{\alpha \beta}\pounds_{v}N^{\mu}_{\alpha \beta}$ with $N^{\mu}_{\alpha \beta}=-\Gamma ^{\mu}_{\alpha \beta}+(1/2)(\Gamma ^{\nu}_{\alpha \nu}\delta ^{\mu}_{\beta}+\Gamma ^{\nu}_{\beta \nu}\delta ^{\mu}_{\alpha})$. To proceed further it is instructive to substitute for $R_{\mu \nu}$ in the above equation in terms of the scalar field and the torsion tensor following \ref{HE_Torsion_02}. With this substitution all the contributions form torsion tensor cancel away and we obtain the following simplified version for the Noether charge density,
\begin{align}
16\pi G~u_{\mu}J^{\mu}[\xi]&=u_{\mu}g^{\alpha \beta}\pounds _{\xi}N^{\mu}_{\alpha \beta}
+16\pi G ~N\left\{\nabla _{\mu}\Phi \nabla _{\nu}\Phi +g_{\mu \nu}V(\Phi)\right\}u^{\mu}u^{\nu}
\label{HE_Torsion_05}
\end{align}
The terms on the right hand side has two contributions --- (a) from the Lie variation of the connection and (b) the projection of the matter energy momentum tensor on the $t=\textrm{constant}$ hypersurface, without any contribution from spacetime torsion whatsoever. In the context of Einstein-Hilbert action the Noether charge density on the left hand side of \ref{HE_Torsion_05} can be written as a total divergence whose integral yields the heat content of the boundary two-surface. However in presence of spacetime torsion such a relation cannot be derived. Furthermore, unlike the situation with the Einstein-Hilbert action, in presence of spacetime torsion the Noether current associated with $v_{\alpha}=\nabla _{\alpha}t$ does not vanish. This has to do with the extra terms present in \ref{NC_Torsion_17} due to the presence of spacetime torsion.  For $v_{\alpha}$, which is gradient of a scalar field, in this case normal to $t=\textrm{constant}$ hypersurface, the associated Noether charge density becomes,
\begin{align}
16\pi G~u_{\mu}J^{\mu}\left[\nabla _{\alpha}t\right]&=-16\pi G~\left(\frac{2}{N}\right)K^{\alpha \sigma}_{~~~\mu}\tau _{\sigma \alpha \nu}u^{\mu}u^{\nu}-\left(\frac{2}{N}\right)\left\{T_{\mu}T_{\nu}-K^{\alpha \beta}_{~~\mu}K_{\beta \alpha \nu}\right\}u^{\mu}u^{\nu}
\label{HE_Torsion_06}
\end{align}
Further, we have the following result connecting the Noether charge density for $\xi ^{\mu}$ and for $\nabla _{\alpha}t$, such that, 
\begin{align}
16\pi G~\left\{u_{\mu}J^{\mu}[\xi]-N^{2}u_{\mu}J^{\mu}[\nabla _{\alpha}t]\right\}=D_{\alpha}(2Na^{\alpha})
\label{HE_Torsion_New_01}
\end{align}
Here $D_{\alpha}$ corresponds to the covariant derivative on the $t=\textrm{constant}$ hypersurface, obtained by projecting the four-dimensional covariant derivative on the three dimensional hypersurface and $a^{\alpha}=u^{\mu}\nabla _{\mu}u^{\alpha}$ is the acceleration associated with $u^{\alpha}$. Since $u_{\mu}=-N\nabla _{\mu}t$, it follows that, the projector will be given by $h^{\mu}_{\nu}=\delta ^{\mu}_{\nu}+u^{\mu}u_{\nu}$. Hence by subtracting \ref{HE_Torsion_06} from \ref{HE_Torsion_05} and using the earlier result on the difference between Noether charge densities, we arrived at,
\begin{align}
16\pi G\Big\{u_{\mu}J^{\mu}[\xi]&-N^{2}u_{\mu}J^{\mu}\left[\nabla _{\alpha}t\right]\Big\}=D_{\alpha}\left(2Na^{\alpha}\right)
\nonumber
\\
&=u_{\mu}g^{\alpha \beta}\pounds _{\xi}N^{\mu}_{\alpha \beta}+16\pi G~N\bar{T}_{\mu \nu}^{\rm (matter)}u^{\mu}u^{\nu}
+16\pi G~N\bar{T}_{\mu \nu}^{\rm (torsion)}u^{\mu}u^{\nu}
\label{HE_Torsion_07}
\end{align}
where the expressions for $\bar{T}_{\mu \nu}^{\rm (matter)}$ and $\bar{T}_{\mu \nu}^{\rm (torsion)}$ from \ref{HE_Torsion_03a} and \ref{HE_Torsion_03b} have been used. Therefore, by rewriting \ref{HE_Torsion_07} in an appropriate manner we finally obtain the following relation,
\begin{align}
u_{\mu}g^{\alpha \beta}\pounds _{\xi}N^{\mu}_{\alpha \beta}&=D_{\alpha}\left(2Na^{\alpha}\right)-16\pi G~N\bar{T}_{\mu \nu}^{\rm (matter)}u^{\mu}u^{\nu}-16\pi G~N\bar{T}_{\mu \nu}^{\rm (torsion)}u^{\mu}u^{\nu}
\label{HE_Torsion_08}
\end{align}
This being one of the key identity which will provide us the desired thermodynamic relation. To write the above in a more suggestive form, it is advantageous to integrate the above equation over a three-dimensional volume $\mathcal{V}$ within the $t=\textrm{constant}$ hypersurface. On the right hand side, the first term will then contribute at the surface $\partial \mathcal{V}$ of $\mathcal{V}$, thanks to the Gauss theorem. This will result into expressions involving $r_{\alpha}a^{\alpha}$, where $r_{\alpha}$ is the normal to $\partial \mathcal{V}$. The above expression makes sense provided the boundary $\partial \mathcal{V}$ is taken to be $N(t,\mathbf{x})=\textrm{constant}$ hypersurface within the $t=\textrm{constant}$ hypersurface, such that $r_{\alpha}=\epsilon D_{\alpha}N(D_{\beta}ND^{\beta}N)^{-1/2}=\epsilon a_{\alpha}/a$. Here $a$ stands for the magnitude of the acceleration four-vector and $\epsilon=+1$ if the normal points outward from the surface and is $-1$ otherwise. Thus finally integration of \ref{HE_Torsion_08} over the volume element $\mathcal{V}$ and subsequent division by $8\pi G$ yields,
\begin{align}
\frac{1}{8\pi G}\int _{\mathcal{V}}d^{3}x\sqrt{h}~u_{\mu}g^{\alpha \beta}\pounds _{\xi}N^{\mu}_{\alpha \beta}&=\frac{\epsilon}{2}\int _{\partial \mathcal{V}} d^{2}x~\left(\frac{Na}{2\pi}\right)
\frac{\sqrt{\sigma}}{G}-\int _{\mathcal{V}}d^{3}x\sqrt{h}~\left\{2N\bar{T}_{\mu \nu}^{\rm (matter)}u^{\mu}u^{\nu}+2N\bar{T}_{\mu \nu}^{\rm (torsion)}u^{\mu}u^{\nu}\right\}
\label{HE_Torsion_09}
\end{align}
Here $\sigma$ is the determinant of the induced metric on the $N=\textrm{constant}$ hypersurface within the $t=\textrm{constant}$ one.  Given the above equation one can define the surface degrees of freedom to be, $\textrm{Area}/G$. Hence even in the presence of spacetime torsion the surface degrees of freedom associated with gravitational dynamics do not change and is still given by the area of the surface. This suggests that as the null limit is taken, for black hole spacetimes as well the entropy-area relation does not get affected by spacetime torsion. Rather torsion affects the bulk degrees of freedom through $\bar{T}_{\mu \nu}^{\rm (torsion)}$.  Thus in presence of spacetime torsion the surface degrees of freedom becomes, 
\begin{align}
N_{\rm sur}=\frac{\textrm{Area}}{G}=\frac{1}{G}\int _{\partial \mathcal{V}}d^{2}x~\sqrt{\sigma}
\label{HE_Torsion_10}
\end{align}
Thus the surface degrees of freedom remains unchanged by the introduction of spacetime torsion and hence  one can conclude the black hole entropy is unaffected by the presence of torsion. Implications of this statement and other avenues to demonstrate the same will be discussed below. Before getting into the bulk degrees of freedom, note that it is advantageous to introduce an average temperature over and above the surface $\partial \mathcal{V}$. In particular, the term $Na/2\pi$ appearing in \ref{HE_Torsion_10} corresponds to the redshifted Davies-Unruh temperature, the locally freely falling observers will associate with the normalized normal $u^{\mu}$. Average of which can be defined as,
\begin{align}
T_{\rm avg}=\frac{1}{\rm Area}\int _{\partial \mathcal{V}} d^{2}x~\sqrt{\sigma}\left(\frac{Na}{2\pi}\right)
\label{HE_Torsion_11}
\end{align}
The bulk degrees of freedom, on the other hand, depends on the matter energy-momentum tensor through $2N\bar{T}_{\mu \nu}^{\rm (matter)}u^{\mu}u^{\nu}$ and on the spacetime torsion through an equivalent of the energy-momentum tensor $\bar{T}_{\mu \nu}^{\rm (torsion)}$ as in \ref{HE_Torsion_03b}. This suggests to introduce the Komar energy density associated with both matter and torsion, such that, 
\begin{align}
\rho _{\rm Komar}^{\rm (matter)}&=2N\bar{T}_{\mu \nu}^{\rm (matter)}u^{\mu}u^{\nu}
\label{HE_Torsion_12a}
\\
\rho_{\rm Komar}^{\rm (torsion)}&=2N\bar{T}_{\mu \nu}^{\rm (torsion)}u^{\mu}u^{\nu}
\label{HE_Torsion_12b}
\end{align}
Thus the bulk degrees of freedom are indeed modified by the introduction of spacetime torsion and hence incorporates contribution from both matter energy density and energy density of the torsion field, such that
\begin{align}
N_{\rm bulk}=\frac{\epsilon}{(1/2)T_{\rm avg}} \int _{\mathcal{V}}d^{3}x\sqrt{h}~\left\{\rho _{\rm Komar}^{\rm (matter)}
+\rho_{\rm Komar}^{\rm (torsion)}\right\}
\label{HE_Torsion_13}
\end{align}
The above definition of bulk degrees of freedom suggests that the bulk region is in equipartition at a temperature $T_{\rm avg}$. Further the factor of $\epsilon$ is included to ensure positivity of the bulk degrees of freedom. For positive Komar energy density one can choose $\epsilon=+1$, while for negative Komar energy density we have $\epsilon=-1$, such that $N_{\rm bulk}$ stays positive. Thus given the definitions of surface and bulk degrees of freedom, presented in \ref{HE_Torsion_10} and \ref{HE_Torsion_13} respectively, along with the expression for average temperature, we can rewrite \ref{HE_Torsion_09}, such that,
\begin{align}
\frac{1}{8\pi G}\int _{\mathcal{V}}d^{3}x\sqrt{h}~u_{\mu}g^{\alpha \beta}\pounds _{\xi}N^{\mu}_{\alpha \beta}
=\frac{\epsilon}{2}T_{\rm avg}\left(N_{\rm sur}-N_{\rm bulk}\right)
\label{HE_Torsion_14}
\end{align}
Thus even when torsion is present, evolution of spacetime originates from the difference between surface and bulk degrees of freedom and structurally coincides with \ref{NC_Review_05}. When the surface and the bulk degrees of freedom coincide, they lead to vanishing of the Lie derivative term and hence $\xi ^{\mu}$ becomes a timelike Killing vector field. This ensures that the spacetime has no dynamics. Thus in the Einstein-Cartan theory as well the departure from Holographic equipartition, or in other words, the difference between suitably defined surface and bulk degrees of freedom is responsible for dynamical evolution of spacetime. This explicitly demonstrates that even in the presence of spacetime torsion, the entropy-area relation, namely $\textrm{entropy}=\textrm{area/4}$ still holds.

An identical conclusion can also be reached in the context of Noether charge as well. In \gr, the total Noether charge within a $N=\textrm{constant}$ hypersurface on the $t=\textrm{constant}$ surface is related to the heat content of the boundary surface. However the above result crucially hinges on the fact that for \gr\ the Noether current associated with $\nabla _{\alpha}t$ identically vanishes. However as evident from \ref{HE_Torsion_06} such is not the situation in the case of Einstein-Cartan theory. Thus in the context of Einstein-Cartan theory we will have the following result associated with the difference between Noether charges with diffeomorphism vector field $\xi^{\mu}$ and $\nabla _{\alpha}t$ respectively, such that,
\begin{align}
\int _{\mathcal{V}} d^{3}x\sqrt{h}\left\{u_{\mu}J^{\mu}[\xi]-N^{2}u_{\mu}J^{\mu}\left[\nabla _{\alpha}t\right]\right\}
=\epsilon \int _{\partial \mathcal{V}}d^{2}x~\left(\frac{Na}{2\pi}\right)\frac{\sqrt{q}}{4G}
=\epsilon \int _{\partial \mathcal{V}}d^{2}x~T_{\rm loc}s~.
\label{HE_Torsion_15}
\end{align}
Here $u_{\mu}J^{\mu}[\xi]$ corresponds to the Noether charge density associated with the vector field $\xi^{\mu}$, while $u_{\mu}J^{\mu}[\nabla t]$ is the Noether charge density associated with the vector field $\nabla _{\mu}t$. In absence of torsion the Noether charge associate with the pure gradient vector field identically vanishes and we recover the result presented in \ref{NC_Review_04}, associated with \gr. While in Einstein-Cartan theory it is the difference between Noether charges that has thermodynamic interpretation. Thus the difference between Noether charges within the $N=\textrm{constant}$ hypersuface associated with $\xi^{\mu}$ and $\nabla _{\alpha}t$ is the heat content of the boundary surface. Note that $T_{\rm loc}=Na/2\pi$ is the redshifted Unruh-Davies temperature associated with the observer with four velocity $u^{\mu}$, who will perceive the local inertial vacuum to be thermally populated. Interestingly, in this context as well the entropy density of the spacetime is given by $\sqrt{q}/4G$, identical to that in general relativity. Thus the Noether charge for Einstein-Cartan gravity leads to the same area-entropy relation. This provides yet another verification of the statement that gravitational entropy is unaffected by the presence of spacetime torsion.
\section{Hamiltonian Analysis in Presence of Torsion and The First Law}
\label{NC_Hamiltonian}

In the previous section we have provided two independent probes to demonstrate that the entropy-area relation is unaffected by the presence of spacetime torsion. Interestingly, it is possible to arrive at the same conclusion starting from the gravitational Hamiltonian in two ways as we will demonstrate now. 

The first and the quickest way to demonstrate that the entropy-area relation remains unaffected is to realize that the incorporation of spacetime torsion does not affect the boundary value problem for general relativity \cite{Parattu:2015gga,Chakraborty:2016yna}. In particular, the boundary term that needs to be added to the gravitational action to make it well-posed, remains the same as the Gibbons-Hawking-York boundary term. Thus the gravitational action in presence of torsion except for the bulk expression given by \ref{NC_Torsion_03} will also have the additional $2K\sqrt{h}$ term integrated over the boundary surface. It is well known that, the $2K\sqrt{h}$ boundary term when integrated over the Rindler horizon in the Euclidean patch results into the entropy-area relation, thanks to the periodicity of the imaginary time coordinate with inverse temperature as the period. In this case as well, since the boundary term is not at all affected by the presence of torsion, it immediately follows that an identical computation on the Rindler horizon will result into the entropy-area relation in this case as well. As an aside we would like to mention that the $(3+1)$ decomposition of the gravitational action as presented in \ref{NC_Torsion_03} except the matter contribution can be performed without much trouble. Since torsion is not dynamical, its conjugate momentum will vanish, while the momentum conjugate to $h_{\mu \nu}$ (the induced metric) remains unchanged. Thus the associated ADM Hamiltonian \cite{Arnowitt:1962hi} will have an identical structure with an additional piece, $2\ntabla_{\alpha}T^{\alpha}+T_{\rho}T^{\rho}+K^{\alpha \rho \mu}K_{\rho \alpha \mu}-16\pi G K^{\alpha}_{~\sigma \beta}\tau ^{\sigma~\beta}_{~\alpha}$, originating from spacetime torsion. This in turn will affect the Hamiltonian constraint by simple addition of the previous term to the corresponding expression for \gr, while the momentum constraint will remain unchanged.

The second approach corresponds to obtaining a first law like structure in the presence of torsion, from which one can read off the associated entropy. For this purpose, we will assume that the spacetimes under consideration is asymptotically flat, which is expected as the effect of torsion must die down at large distance from the source. Further, we will make a slight change in the nomenclature as well, we will assume that we are working in a spacetime which admits both timelike ($t^{\mu}$) and spacelike ($\phi_{(a)} ^{\mu}$) Killing vector fields, whose appropriate combination $\xi^{\mu}_{\rm K}$ is the Killing vector field whose norm vanishes on the horizon. The subscript $\textrm{K}$ to $\xi^{\mu}_{\rm K}$ is to distinguish it from the vector $\xi^{\mu}$ we have introduced in \ref{NC_Review}.  

With this preamble, let us compute the variation of the symplectic potential, dubbed as the Hamiltonian. Thus given the Noether current $J^{\mu}[\xi_{\rm K}]$ and the boundary term originating from variation of the action, we can write down variation of the Hamiltonian using the symplectic structure in the following manner \cite{Iyer:1994ys},
\begin{align}\label{H_Torsion_01}
\delta H= \delta \int _{\mathcal{C}} d^{3}x\sqrt{h}~u_{\mu}J^{\mu}[\xi_{\rm K}]
-\int _{\partial \mathcal{C}_{\infty}}d^{2}x\sqrt{q}~r^{\alpha \beta}\xi ^{\mu}_{\rm K}\epsilon _{\alpha \beta \mu \nu}\delta V^{\nu}
\end{align}
Here the quantity $\delta V^{\nu}$ is defined in \ref{NC_Torsion_08}, $r^{\alpha \beta}$ is the bi-normal to the two-surface $\partial \mathcal{C}_{\infty}$ and $\epsilon_{\alpha \beta \mu \nu}$ is the Levi-Civita symbol. Moreover, $\mathcal{C}$ depicts a Cauchy surface in an asymptotically flat spacetime, such that it inherits a single boundary $\partial \mathcal{C}_{\infty}$ at infinity, while the inner horizon is assumed to be compact.  For the existence of a Hamiltonian it is necessary that the term on the boundary $\partial \mathcal{C}_{\infty}$ be written as 
\begin{align}\label{H_Torsion_02}
\int _{\partial \mathcal{C}_{\infty}}d^{2}x\sqrt{q}~r^{\alpha \beta}\xi ^{\mu}_{\rm K}\epsilon _{\alpha \beta \mu \nu}\delta V^{\nu}
=\delta \int _{\partial \mathcal{C}_{\infty}}d^{2}x\sqrt{q}~r^{\alpha \beta}\xi ^{\mu}_{\rm K}\epsilon _{\alpha \beta \mu \nu} B^{\nu}
\end{align}
where the vector $B^{\mu}$ needs to be determined.
 
Thus one can identify the Hamiltonian as inheriting contributions from both $J^{\mu}[\xi_{\rm K}]$ as well as from the boundary term $B^{\mu}$. Further the Noether current as presented in \ref{NC_Torsion_N1} has two parts, one originating from general relativity ($J^{\mu}_{\rm gr}[v]$), while the other depending exclusively on spacetime torsion ($J^{\mu}_{\rm tor}[v]$). Interestingly the general relativity contribution can be converted to a surface integral as one can have a Noether potential for the same. Thus the Hamiltonian can be written as, 
\begin{align}\label{H_Torsion_03}
H=\int _{\partial \mathcal{C}_{\infty}}d^{2}x\sqrt{q}~r_{\alpha \beta}\left(J^{\alpha \beta}_{\rm gr}[\xi_{\rm K}]-\epsilon ^{\alpha \beta}_{~~~\mu \nu}\xi ^{\mu}_{\rm K}B^{\nu}\right)
+\int _{\mathcal{C}}d^{3}x\sqrt{h}~u_{\mu}J^{\mu}_{\rm tor}[\xi_{\rm K}]
\end{align}
Since the boundary contribution as in \ref{NC_Torsion_08} as well as $J^{\alpha \beta}_{\rm gr}[\xi_{\rm K}]$ does not involve any torsional degrees of freedom, it follows that the first term is identical to that in general relativity, while the second term, which is a volume term encodes the torsional degrees of freedom. Another case of interest corresponds to the situation in which the Killing vector field $\xi^{\mu}_{\rm K}$ generates the asymptotic time translation, i.e, $\xi^{\mu}_{\rm K}\rightarrow t^{\mu}$. In this case the value of the Hamiltonian defines the energy associated with the system. In particular, we will have,
\begin{align}\label{H_Torsion_04}
\mathcal{E}&=\int _{\partial \mathcal{C}_{\infty}}d^{2}x\sqrt{q}~r_{\alpha \beta}\left(J^{\alpha \beta}_{\rm gr}[t]-\epsilon ^{\alpha \beta}_{~~~\mu \nu}\xi ^{\mu}_{\rm K}B^{\nu}\right)
+\int _{\mathcal{C}}d^{3}x\sqrt{h}~u_{\mu}J^{\mu}_{\rm tor}[t]
\nonumber
\\
&=\int _{\partial \mathcal{C}_{\infty}}d^{2}x\sqrt{q}~r_{\alpha \beta}\left(J^{\alpha \beta}_{\rm gr}[t]-\epsilon ^{\alpha \beta}_{~~~\mu \nu}\xi ^{\mu}_{\rm K}B^{\nu}\right)
+\mathcal{E}_{\rm torsion}
\end{align}
For asymptotically flat spacetimes the first integral is essentially an integral over the two-surface at infinity and will coincide with the ADM mass of the vacuum spacetime. Since torsion identically vanishes in absence of matter, it follows that the volume term will also not contribute. Thus for vacuum spacetimes the energy defined above coincides with the ADM mass as it should. On the other hand in presence of matter, the spacetime torsion will be non-zero and hence the effect from torsion on energy $\mathcal{E}$ will be encoded in the volume integral of \ref{H_Torsion_04}. 

Again going back to the general case with $\xi^{\mu}_{\rm K}$ as the diffeomorphism vector field, one may be able to write the same as a combination of time translation Killing field $t^{\mu}$ and the rotational Killing field(s) $\phi ^{\mu}_{(a)}$, by introducing some appropriate linear combinations with constant coefficients, such that,
\begin{align}\label{H_Torsion_05}
\xi ^{\mu}_{\rm K}=t^{\mu}+\Omega _{\rm H}^{(a)}\phi ^{\mu}_{(a)}
\end{align}
Here $\Omega _{\rm H}^{(a)}$ are the constant coefficients. In presence of Killing horizon in an asymptotically flat spacetime the variation of the Noether charge over an equal time hypersurface bounded by event horizon $\mathcal{H}$ and two-surface at infinity $\partial \mathcal{V}_{\infty}$ can be written as,
\begin{align}\label{H_Torsion_06}
\delta \int _{\mathcal{H}}d^{2}x\sqrt{q}~r_{\alpha \beta}J^{\alpha \beta}_{\rm gr}[\xi_{\rm K}]
&=\delta \int _{\partial \mathcal{V}_{\infty}}d^{2}x\sqrt{q}~r_{\alpha \beta}J^{\alpha \beta}_{\rm gr}[\xi_{\rm K}]
\nonumber
\\
&-\delta \int _{\partial \mathcal{V}_{\infty}}d^{2}x\sqrt{q}~r_{\alpha \beta}\epsilon ^{\alpha \beta}_{~~~\mu \nu}\xi ^{\mu}_{\rm K}B^{\nu}[\xi_{\rm K}]
+\delta \int _{\mathcal{V}}d^{3}x\sqrt{h}~u_{\mu}J^{\mu}_{\rm tor}[\xi_{\rm K}]
\end{align}
where we have used the result that the Killing field $\xi^{\mu}_{\rm K}$ vanishes on the event horizon $\mathcal{H}$. Since the hypersurface $\mathcal{V}$ is also in an asymptotically flat spacetime it must match with the Cauchy surface except for the interior, such that $\partial \mathcal{V}_{\infty}=\partial \mathcal{C}_{\infty}$. As evident from \ref{H_Torsion_05}, the Killing field can be linearly decomposed and the above variation of the Noether charge presented in \ref{H_Torsion_06} can be written as,
\begin{align}\label{H_Torsion_07}
\delta \int _{\mathcal{H}}d^{2}x\sqrt{q}~r_{\alpha \beta}J^{\alpha \beta}_{\rm gr}[\xi_{\rm K}]
&=\delta \int _{\partial \mathcal{C}_{\infty}}d^{2}x\sqrt{q}~r_{\alpha \beta}J^{\alpha \beta}_{\rm gr}[t]
-\delta \int _{\partial \mathcal{C}_{\infty}}d^{2}x\sqrt{q}~r_{\alpha \beta}\epsilon ^{\alpha \beta}_{~~~\mu \nu}t^{\mu}B^{\nu}[t]
+\delta \int _{\mathcal{C}}d^{3}x\sqrt{h}~u_{\mu}J^{\mu}_{\rm tor}[t]
\nonumber
\\
&+\Omega _{\rm H}^{(a)}\left\{\delta \int _{\partial \mathcal{C}_{\infty}}d^{2}x\sqrt{q}~r_{\alpha \beta}J^{\alpha \beta}_{\rm gr}[\phi_{(a)}]
+\delta \int _{\mathcal{C}}d^{3}x\sqrt{h}~u_{\mu}J^{\mu}_{\rm tor}[\phi_{(a)}] \right\}
\end{align}
Note that the term involving  $\phi^{\mu}_{(a)}B^{\nu}$ on the boundary is absent due to antisymmetry of the Levi-Civita tensor. Collecting all the terms on the right hand side, it is clear that they can be separated into two parts --- (a) one due to the time translation field, which corresponds to the first three terms on the right hand side and is equal to $\delta \mathcal{E}$ as evident from \ref{H_Torsion_04}; (b) The two terms in the second line is connected to the rotational Killing field $\phi ^{\mu}_{(a)}$ and corresponds to the total Noether charge associated with $\phi ^{\mu}_{(a)}$ and hence can be defined as the negative of the angular momentum $\mathcal{J}_{(a)}$ associated with the spacetime \cite{Iyer:1994ys}. In vacuum situations the torsional piece identically vanishes and we end up getting the standard result for general relativity. Thus alike the definition of energy, the definition of angular momentum as well gets modified in presence of torsion. Finally incorporating all these contributions, one can write down \ref{H_Torsion_07} in a compact form, such that,
\begin{align}\label{H_Torsion_08}
\int _{\rm \mathcal{H}}d^{2}x\sqrt{q}~r_{\alpha \beta}J^{\alpha \beta}_{\rm gr}[\xi_{\rm K}]
=\frac{\kappa}{2\pi}\delta \left(\frac{\textrm{Area}}{4G}\right)
=\delta \mathcal{E}-\Omega _{\rm H}^{(a)}\delta \mathcal{J}_{(a)}~.
\end{align}
As emphasized earlier, here $\mathcal{J}_{(a)}$ is the negative of the Noether charge associated with rotational Killing vector field at infinity along with a volume term due to torsion and $\mathcal{E}$ is given by \ref{H_Torsion_04}, both of which involves the effect from torsion. This is essentially the first law in a spacetime inheriting torsion. Intriguingly, the left hand side of \ref{H_Torsion_08} originates from \gr\ alone and is equal to $T\delta S$, where $S=\textrm{Area}/4G$. Thus the above derivation of first law for spacetimes with torsion provides yet another verification of the entropy-area relation in the context of Einstein-Cartan theory. In conformity with the earlier findings, in the context of first law as well torsion only affects the expression for bulk quantities, e.g., $\mathcal{E}$ and $\mathcal{J}$ respectively. This further bolsters our claim regarding black hole entropy in presence of torsion presented in this work. 
\section{Concluding Remarks}
\label{NC_Conclusion}

In this work we have explicitly demonstrated, that even though the presence of spacetime torsion modifies the gravitational Lagrangian and hence the field equations in a \emph{non-trivial} manner, there is \emph{no} effect of the same on the black hole entropy. In other words, even in the presence of torsion, the entropy-area relation holds identically. We have arrived at this conclusion following several routes:
\begin{itemize}

\item We have shown that even in presence of spacetime torsion, the evolution of the spacetime is governed by the difference between suitably defined bulk and surface degrees of freedom. Here the surface degrees of freedom inherits no contribution from spacetime torsion and is proportional to area.

\item The difference between Noether charges for suitable diffeomorphism vector fields is equal to the heat content of the boundary surface. Here also the boundary entropy is proportional to the area of the boundary surface.

\item The Gibbons-Hawking-York boundary term remains unchanged in presence of torsion and hence yields the entropy-area relation, when evaluated in the near horizon regime using Euclidean methods.

\item Finally, variation of the Noether charge over the black hole horizon can be written in a first law like form, where the entropy is still given by (area/4), while the energy and angular momentum changes due to presence of torsion.

\end{itemize}

It is also possible to argue the same starting from the $(3+1)$ decomposition of the gravitational Lagrangian, resulting into the ADM-like Hamiltonian with the boundary contribution being completely independent of torsion. Thus we observe that black hole entropy in presence of torsion is indeed proportional to area, while the bulk contribution will indeed inherit the effect from torsion. This shows that any \emph{non-trivial} modifications of the gravitational action does \emph{not} necessarily modify the black hole entropy. This result has significant physical as well as philosophical fallout. For example, the above study lends into several important questions that one may wish to answer, viz., what happens to black hole entropy in Lovelock theories of gravity, \emph{but} in the presence of spacetime torsion. Does it change from the Wald entropy or not? In other words, does the above peculiar result associated with Einstein-Cartan theory transcends general relativity? Also in the context of AdS black holes in Einstein-Cartan theory, our result suggests that the torsional degrees of freedom does not contribute to black hole entropy and hence it is worth asking whether the equivalent CFT description of Einstein-Cartan theory in AdS spacetime is identical to general relativity or not. These are some of the question worth wondering about and they shows the importance of spacetime torsion in the understanding of microscopic structure of spacetime in a better and systematic manner. Our result may be a first and primitive step towards that direction. 
\section*{Acknowledgements}

Research of SC is funded by the INSPIRE Faculty Fellowship (Reg. No. DST/INSPIRE/04/2018/000893) from Department of Science and Technology, Government of India. 
\bibliography{References}

\providecommand{\href}[2]{#2}\begingroup\raggedright\begin{thebibliography}{10}

\bibitem{Bardeen1973}
J.~M. Bardeen, B.~Carter, and S.~W. Hawking, ``The four laws of black hole
  mechanics,'' \href{http://dx.doi.org/10.1007/BF01645742}{{\em Communications
  in Mathematical Physics} {\bfseries 31} no.~2, (Jun, 1973) 161--170}.
  \url{https://doi.org/10.1007/BF01645742}.

\bibitem{Bekenstein:1973ur}
J.~D. Bekenstein, ``{Black holes and entropy},''
\href{http://dx.doi.org/10.1103/PhysRevD.7.2333}{{\em Phys. Rev.} {\bfseries
  D7} (1973) 2333--2346}.

\bibitem{Hawking:1974sw}
S.~W. Hawking, ``{Particle Creation by Black Holes},''
  \href{http://dx.doi.org/10.1007/BF02345020, 10.1007/BF01608497}{{\em Commun.
  Math. Phys.} {\bfseries 43} (1975) 199--220}.
[erratum, ibid 167(1975)].

\bibitem{Gibbons:1976ue}
G.~W. Gibbons and S.~W. Hawking, ``{Action Integrals and Partition Functions in
  Quantum Gravity},''
\href{http://dx.doi.org/10.1103/PhysRevD.15.2752}{{\em Phys. Rev.} {\bfseries
  D15} (1977) 2752--2756}.

\bibitem{Unruh:1976db}
W.~G. Unruh, ``{Notes on black hole evaporation},''
\href{http://dx.doi.org/10.1103/PhysRevD.14.870}{{\em Phys. Rev.} {\bfseries
  D14} (1976) 870}.

\bibitem{THOOFT1985727}
G.~'t~Hooft, ``On the quantum structure of a black hole,''
  \href{http://dx.doi.org/http://dx.doi.org/10.1016/0550-3213(85)90418-3}{{\em
  Nuclear Physics B} {\bfseries 256} (1985) 727 -- 745}.
  \url{http://www.sciencedirect.com/science/article/pii/0550321385904183}.

\bibitem{Susskind:1993ws}
L.~Susskind, ``{Some speculations about black hole entropy in string theory},''
\href{http://arxiv.org/abs/hep-th/9309145}{{\ttfamily arXiv:hep-th/9309145
  [hep-th]}}.

\bibitem{Sen:1995in}
A.~Sen, ``{Extremal black holes and elementary string states},''
  \href{http://dx.doi.org/10.1142/S0217732395002234}{{\em Mod. Phys. Lett.}
  {\bfseries A10} (1995) 2081--2094},
\href{http://arxiv.org/abs/hep-th/9504147}{{\ttfamily arXiv:hep-th/9504147
  [hep-th]}}.

\bibitem{Horowitz:1996qd}
G.~T. Horowitz, ``{The Origin of black hole entropy in string theory},''
  \href{http://arxiv.org/abs/gr-qc/9604051}{{\ttfamily arXiv:gr-qc/9604051
  [gr-qc]}}.
[Astrophys. Space Sci. Libr.211,46(1997)].

\bibitem{Strominger:1996sh}
A.~Strominger and C.~Vafa, ``{Microscopic origin of the Bekenstein-Hawking
  entropy},'' \href{http://dx.doi.org/10.1016/0370-2693(96)00345-0}{{\em Phys.
  Lett.} {\bfseries B379} (1996) 99--104},
\href{http://arxiv.org/abs/hep-th/9601029}{{\ttfamily arXiv:hep-th/9601029
  [hep-th]}}.

\bibitem{Rovelli:1996dv}
C.~Rovelli, ``{Black hole entropy from loop quantum gravity},''
  \href{http://dx.doi.org/10.1103/PhysRevLett.77.3288}{{\em Phys. Rev. Lett.}
  {\bfseries 77} (1996) 3288--3291},
\href{http://arxiv.org/abs/gr-qc/9603063}{{\ttfamily arXiv:gr-qc/9603063
  [gr-qc]}}.

\bibitem{Bombelli:1986rw}
L.~Bombelli, R.~K. Koul, J.~Lee, and R.~D. Sorkin, ``{A Quantum Source of
  Entropy for Black Holes},''
\href{http://dx.doi.org/10.1103/PhysRevD.34.373}{{\em Phys. Rev.} {\bfseries
  D34} (1986) 373--383}.

\bibitem{Jacobson:2005kr}
T.~Jacobson, D.~Marolf, and C.~Rovelli, ``{Black hole entropy: Inside or
  out?},'' \href{http://dx.doi.org/10.1007/s10773-005-8896-z}{{\em Int. J.
  Theor. Phys.} {\bfseries 44} (2005) 1807--1837},
\href{http://arxiv.org/abs/hep-th/0501103}{{\ttfamily arXiv:hep-th/0501103
  [hep-th]}}.

\bibitem{Jacobson:2003wv}
T.~Jacobson and R.~Parentani, ``{Horizon entropy},''
  \href{http://dx.doi.org/10.1023/A:1023785123428}{{\em Found. Phys.}
  {\bfseries 33} (2003) 323--348},
\href{http://arxiv.org/abs/gr-qc/0302099}{{\ttfamily arXiv:gr-qc/0302099
  [gr-qc]}}.

\bibitem{Callan:1995hn}
C.~G. Callan, J.~M. Maldacena, and A.~W. Peet, ``{Extremal black holes as
  fundamental strings},''
  \href{http://dx.doi.org/10.1016/0550-3213(96)00315-X}{{\em Nucl. Phys.}
  {\bfseries B475} (1996) 645--678},
\href{http://arxiv.org/abs/hep-th/9510134}{{\ttfamily arXiv:hep-th/9510134
  [hep-th]}}.

\bibitem{Duff:1994an}
M.~J. Duff, R.~R. Khuri, and J.~X. Lu, ``{String solitons},''
  \href{http://dx.doi.org/10.1016/0370-1573(95)00002-X}{{\em Phys. Rept.}
  {\bfseries 259} (1995) 213--326},
\href{http://arxiv.org/abs/hep-th/9412184}{{\ttfamily arXiv:hep-th/9412184
  [hep-th]}}.

\bibitem{Larsen:1995ss}
F.~Larsen and F.~Wilczek, ``{Internal structure of black holes},''
  \href{http://dx.doi.org/10.1016/0370-2693(96)00220-1}{{\em Phys. Lett.}
  {\bfseries B375} (1996) 37--42},
\href{http://arxiv.org/abs/hep-th/9511064}{{\ttfamily arXiv:hep-th/9511064
  [hep-th]}}.

\bibitem{Carlip:1995cd}
S.~Carlip, ``{Statistical mechanics and black hole entropy},''
\href{http://arxiv.org/abs/gr-qc/9509024}{{\ttfamily arXiv:gr-qc/9509024
  [gr-qc]}}.

\bibitem{Ashtekar:1997yu}
A.~Ashtekar, J.~Baez, A.~Corichi, and K.~Krasnov, ``{Quantum geometry and black
  hole entropy},'' \href{http://dx.doi.org/10.1103/PhysRevLett.80.904}{{\em
  Phys. Rev. Lett.} {\bfseries 80} (1998) 904--907},
\href{http://arxiv.org/abs/gr-qc/9710007}{{\ttfamily arXiv:gr-qc/9710007
  [gr-qc]}}.

\bibitem{Date:2008rb}
G.~Date, R.~K. Kaul, and S.~Sengupta, ``{Topological Interpretation of
  Barbero-Immirzi Parameter},''
  \href{http://dx.doi.org/10.1103/PhysRevD.79.044008}{{\em Phys. Rev.}
  {\bfseries D79} (2009) 044008},
\href{http://arxiv.org/abs/0811.4496}{{\ttfamily arXiv:0811.4496 [gr-qc]}}.

\bibitem{Frodden:2012dq}
E.~Frodden, M.~Geiller, K.~Noui, and A.~Perez, ``{Black Hole Entropy from
  complex Ashtekar variables},''
  \href{http://dx.doi.org/10.1209/0295-5075/107/10005}{{\em EPL} {\bfseries
  107} no.~1, (2014) 10005},
\href{http://arxiv.org/abs/1212.4060}{{\ttfamily arXiv:1212.4060 [gr-qc]}}.

\bibitem{Pranzetti:2014tla}
D.~Pranzetti and H.~Sahlmann, ``{Horizon entropy with loop quantum gravity
  methods},'' \href{http://dx.doi.org/10.1016/j.physletb.2015.04.070}{{\em
  Phys. Lett.} {\bfseries B746} (2015) 209--216},
\href{http://arxiv.org/abs/1412.7435}{{\ttfamily arXiv:1412.7435 [gr-qc]}}.

\bibitem{Neiman:2013ap}
Y.~Neiman, ``{The imaginary part of the gravity action and black hole
  entropy},'' \href{http://dx.doi.org/10.1007/JHEP04(2013)071}{{\em JHEP}
  {\bfseries 04} (2013) 071},
\href{http://arxiv.org/abs/1301.7041}{{\ttfamily arXiv:1301.7041 [gr-qc]}}.

\bibitem{Jacobson:1995ab}
T.~Jacobson, ``{Thermodynamics of space-time: The Einstein equation of
  state},'' \href{http://dx.doi.org/10.1103/PhysRevLett.75.1260}{{\em Phys.
  Rev. Lett.} {\bfseries 75} (1995) 1260--1263},
\href{http://arxiv.org/abs/gr-qc/9504004}{{\ttfamily arXiv:gr-qc/9504004
  [gr-qc]}}.

\bibitem{Padmanabhan:2002sha}
T.~Padmanabhan, ``{Classical and quantum thermodynamics of horizons in
  spherically symmetric space-times},''
  \href{http://dx.doi.org/10.1088/0264-9381/19/21/306}{{\em Class. Quant.
  Grav.} {\bfseries 19} (2002) 5387--5408},
\href{http://arxiv.org/abs/gr-qc/0204019}{{\ttfamily arXiv:gr-qc/0204019
  [gr-qc]}}.

\bibitem{Padmanabhan:2003gd}
T.~Padmanabhan, ``{Gravity and the thermodynamics of horizons},''
  \href{http://dx.doi.org/10.1016/j.physrep.2004.10.003}{{\em Phys. Rept.}
  {\bfseries 406} (2005) 49--125},
\href{http://arxiv.org/abs/gr-qc/0311036}{{\ttfamily arXiv:gr-qc/0311036
  [gr-qc]}}.

\bibitem{Padmanabhan:2009vy}
T.~Padmanabhan, ``{Thermodynamical Aspects of Gravity: New insights},''
  \href{http://dx.doi.org/10.1088/0034-4885/73/4/046901}{{\em Rept. Prog.
  Phys.} {\bfseries 73} (2010) 046901},
\href{http://arxiv.org/abs/0911.5004}{{\ttfamily arXiv:0911.5004 [gr-qc]}}.

\bibitem{Kothawala:2009kc}
D.~Kothawala and T.~Padmanabhan, ``{Thermodynamic structure of Lanczos-Lovelock
  field equations from near-horizon symmetries},''
  \href{http://dx.doi.org/10.1103/PhysRevD.79.104020}{{\em Phys. Rev.}
  {\bfseries D79} (2009) 104020},
\href{http://arxiv.org/abs/0904.0215}{{\ttfamily arXiv:0904.0215 [gr-qc]}}.

\bibitem{Chakraborty:2015aja}
S.~Chakraborty, K.~Parattu, and T.~Padmanabhan, ``{Gravitational field
  equations near an arbitrary null surface expressed as a thermodynamic
  identity},'' \href{http://dx.doi.org/10.1007/JHEP10(2015)097}{{\em JHEP}
  {\bfseries 10} (2015) 097},
\href{http://arxiv.org/abs/1505.05297}{{\ttfamily arXiv:1505.05297 [gr-qc]}}.

\bibitem{Chakraborty:2015wma}
S.~Chakraborty, ``{Lanczos-Lovelock gravity from a thermodynamic
  perspective},'' \href{http://dx.doi.org/10.1007/JHEP08(2015)029}{{\em JHEP}
  {\bfseries 08} (2015) 029},
\href{http://arxiv.org/abs/1505.07272}{{\ttfamily arXiv:1505.07272 [gr-qc]}}.

\bibitem{Dey:2016zka}
R.~Dey, S.~Liberati, and A.~Mohd, ``{Higher derivative gravity: field equation
  as the equation of state},''
  \href{http://dx.doi.org/10.1103/PhysRevD.94.044013}{{\em Phys. Rev.}
  {\bfseries D94} no.~4, (2016) 044013},
\href{http://arxiv.org/abs/1605.04789}{{\ttfamily arXiv:1605.04789 [gr-qc]}}.

\bibitem{Sarkar:2013swa}
S.~Sarkar and A.~C. Wall, ``{Generalized second law at linear order for actions
  that are functions of Lovelock densities},''
  \href{http://dx.doi.org/10.1103/PhysRevD.88.044017}{{\em Phys. Rev.}
  {\bfseries D88} (2013) 044017},
\href{http://arxiv.org/abs/1306.1623}{{\ttfamily arXiv:1306.1623 [gr-qc]}}.

\bibitem{Eling:2006aw}
C.~Eling, R.~Guedens, and T.~Jacobson, ``{Non-equilibrium thermodynamics of
  spacetime},'' \href{http://dx.doi.org/10.1103/PhysRevLett.96.121301}{{\em
  Phys. Rev. Lett.} {\bfseries 96} (2006) 121301},
\href{http://arxiv.org/abs/gr-qc/0602001}{{\ttfamily arXiv:gr-qc/0602001
  [gr-qc]}}.

\bibitem{Wald:1993nt}
R.~M. Wald, ``{Black hole entropy is the Noether charge},''
  \href{http://dx.doi.org/10.1103/PhysRevD.48.R3427}{{\em Phys.Rev.} {\bfseries
  D48} (1993) 3427--3431},
\href{http://arxiv.org/abs/gr-qc/9307038}{{\ttfamily arXiv:gr-qc/9307038
  [gr-qc]}}.

\bibitem{Iyer:1994ys}
V.~Iyer and R.~M. Wald, ``{Some properties of Noether charge and a proposal for
  dynamical black hole entropy},''
  \href{http://dx.doi.org/10.1103/PhysRevD.50.846}{{\em Phys. Rev.} {\bfseries
  D50} (1994) 846--864},
\href{http://arxiv.org/abs/gr-qc/9403028}{{\ttfamily arXiv:gr-qc/9403028
  [gr-qc]}}.

\bibitem{Jacobson:1993xs}
T.~Jacobson and R.~C. Myers, ``{Black hole entropy and higher curvature
  interactions},'' \href{http://dx.doi.org/10.1103/PhysRevLett.70.3684}{{\em
  Phys.Rev.Lett.} {\bfseries 70} (1993) 3684--3687},
\href{http://arxiv.org/abs/hep-th/9305016}{{\ttfamily arXiv:hep-th/9305016
  [hep-th]}}.

\bibitem{Sheykhi:2007gi}
A.~Sheykhi, B.~Wang, and R.-G. Cai, ``{Deep Connection Between Thermodynamics
  and Gravity in Gauss-Bonnet Braneworld},''
  \href{http://dx.doi.org/10.1103/PhysRevD.76.023515}{{\em Phys. Rev.}
  {\bfseries D76} (2007) 023515},
\href{http://arxiv.org/abs/hep-th/0701261}{{\ttfamily arXiv:hep-th/0701261
  [hep-th]}}.

\bibitem{Wu:2008ir}
S.-F. Wu, B.~Wang, G.-H. Yang, and P.-M. Zhang, ``{The Generalized second law
  of thermodynamics in generalized gravity theories},''
  \href{http://dx.doi.org/10.1088/0264-9381/25/23/235018}{{\em Class. Quant.
  Grav.} {\bfseries 25} (2008) 235018},
\href{http://arxiv.org/abs/0801.2688}{{\ttfamily arXiv:0801.2688 [hep-th]}}.

\bibitem{Kothawala:2007em}
D.~Kothawala, S.~Sarkar, and T.~Padmanabhan, ``{Einstein's equations as a
  thermodynamic identity: The Cases of stationary axisymmetric horizons and
  evolving spherically symmetric horizons},''
  \href{http://dx.doi.org/10.1016/j.physletb.2007.07.021}{{\em Phys. Lett.}
  {\bfseries B652} (2007) 338--342},
\href{http://arxiv.org/abs/gr-qc/0701002}{{\ttfamily arXiv:gr-qc/0701002
  [gr-qc]}}.

\bibitem{Chakraborty:2010wn}
S.~Chakraborty, R.~Biswas, and N.~Mazumder, ``{Unified First Law and Some
  Comments},'' \href{http://dx.doi.org/10.1393/ncb/i2010-10912-5}{{\em Nuovo
  Cim.} {\bfseries B125} (2011) 1209--1214},
\href{http://arxiv.org/abs/1006.1169}{{\ttfamily arXiv:1006.1169 [gr-qc]}}.

\bibitem{Paranjape:2006ca}
A.~Paranjape, S.~Sarkar, and T.~Padmanabhan, ``{Thermodynamic route to field
  equations in Lancos-Lovelock gravity},''
  \href{http://dx.doi.org/10.1103/PhysRevD.74.104015}{{\em Phys. Rev.}
  {\bfseries D74} (2006) 104015},
\href{http://arxiv.org/abs/hep-th/0607240}{{\ttfamily arXiv:hep-th/0607240
  [hep-th]}}.

\bibitem{Chakraborty:2014rga}
S.~Chakraborty and T.~Padmanabhan, ``{Evolution of Spacetime arises due to the
  departure from Holographic Equipartition in all Lanczos-Lovelock Theories of
  Gravity},'' \href{http://dx.doi.org/10.1103/PhysRevD.90.124017}{{\em Phys.
  Rev.} {\bfseries D90} no.~12, (2014) 124017},
\href{http://arxiv.org/abs/1408.4679}{{\ttfamily arXiv:1408.4679 [gr-qc]}}.

\bibitem{Liberati:2017vse}
C.~Pacilio and S.~Liberati, ``{First law of black holes with a universal
  horizon},'' \href{http://dx.doi.org/10.1103/PhysRevD.96.104060}{{\em Phys.
  Rev.} {\bfseries D96} no.~10, (2017) 104060},
\href{http://arxiv.org/abs/1709.05802}{{\ttfamily arXiv:1709.05802 [gr-qc]}}.

\bibitem{DePaoli:2018erh}
E.~De~Paoli and S.~Speziale, ``{A gauge-invariant symplectic potential for
  tetrad general relativity},''
\href{http://arxiv.org/abs/1804.09685}{{\ttfamily arXiv:1804.09685 [gr-qc]}}.

\bibitem{Prabhu:2015vua}
K.~Prabhu, ``{The First Law of Black Hole Mechanics for Fields with Internal
  Gauge Freedom},'' \href{http://dx.doi.org/10.1088/1361-6382/aa536b}{{\em
  Class. Quant. Grav.} {\bfseries 34} no.~3, (2017) 035011},
\href{http://arxiv.org/abs/1511.00388}{{\ttfamily arXiv:1511.00388 [gr-qc]}}.

\bibitem{Barnich:2016rwk}
G.~Barnich, P.~Mao, and R.~Ruzziconi, ``{Conserved currents in the Cartan
  formulation of general relativity},'' in {\em {About Various Kinds of
  Interactions: Workshop in honour of Professor Philippe Spindel Mons, Belgium,
  June 4-5, 2015}}.
\newblock 2016.
\newblock \href{http://arxiv.org/abs/1611.01777}{{\ttfamily arXiv:1611.01777
  [gr-qc]}}.
\newblock
\url{https://inspirehep.net/record/1495975/files/arXiv:1611.01777.pdf}.
\newblock

\bibitem{Blagojevic:2006jk}
M.~Blagojevic and B.~Cvetkovic, ``{Black hole entropy in 3-D gravity with
  torsion},'' \href{http://dx.doi.org/10.1088/0264-9381/23/14/013}{{\em Class.
  Quant. Grav.} {\bfseries 23} (2006) 4781},
\href{http://arxiv.org/abs/gr-qc/0601006}{{\ttfamily arXiv:gr-qc/0601006
  [gr-qc]}}.

\bibitem{Dey:2017fld}
R.~Dey, S.~Liberati, and D.~Pranzetti, ``{Spacetime thermodynamics in the
  presence of torsion},''
  \href{http://dx.doi.org/10.1103/PhysRevD.96.124032}{{\em Phys. Rev.}
  {\bfseries D96} no.~12, (2017) 124032},
\href{http://arxiv.org/abs/1709.04031}{{\ttfamily arXiv:1709.04031 [gr-qc]}}.

\bibitem{Padmanabhan:2013nxa}
T.~Padmanabhan, ``{General Relativity from a Thermodynamic Perspective},''
  \href{http://dx.doi.org/10.1007/s10714-014-1673-7}{{\em Gen. Rel. Grav.}
  {\bfseries 46} (2014) 1673},
\href{http://arxiv.org/abs/1312.3253}{{\ttfamily arXiv:1312.3253 [gr-qc]}}.

\bibitem{Chakraborty:2015hna}
S.~Chakraborty and T.~Padmanabhan, ``{Thermodynamical interpretation of the
  geometrical variables associated with null surfaces},''
  \href{http://dx.doi.org/10.1103/PhysRevD.92.104011}{{\em Phys. Rev.}
  {\bfseries D92} no.~10, (2015) 104011},
\href{http://arxiv.org/abs/1508.04060}{{\ttfamily arXiv:1508.04060 [gr-qc]}}.

\bibitem{Hehl:1976kj}
F.~W. Hehl, P.~Von Der~Heyde, G.~D. Kerlick, and J.~M. Nester, ``{General
  Relativity with Spin and Torsion: Foundations and Prospects},''
\href{http://dx.doi.org/10.1103/RevModPhys.48.393}{{\em Rev. Mod. Phys.}
  {\bfseries 48} (1976) 393--416}.

\bibitem{Trautman:2006fp}
A.~Trautman, ``{Einstein-Cartan theory},''
\href{http://arxiv.org/abs/gr-qc/0606062}{{\ttfamily arXiv:gr-qc/0606062
  [gr-qc]}}.

\bibitem{Nester:2012ib}
J.~M. Nester and C.-H. Wang, ``{Can torsion be treated as just another tensor
  field?},''
\href{http://dx.doi.org/10.1142/S2010194512004229}{{\em Int. J. Mod. Phys.
  Conf. Ser.} {\bfseries 07} (2012) 158}.

\bibitem{Chen:2015vya}
C.-M. Chen, J.~M. Nester, and R.-S. Tung, ``{Gravitational energy for GR and
  Poincaré gauge theories: A covariant Hamiltonian approach},''
  \href{http://dx.doi.org/10.1142/S0218271815300268}{{\em Int. J. Mod. Phys.}
  {\bfseries D24} no.~11, (2015) 1530026},
\href{http://arxiv.org/abs/1507.07300}{{\ttfamily arXiv:1507.07300 [gr-qc]}}.

\bibitem{Nester:2016rsy}
J.~M. Nester and C.-M. Chen, ``{Gravity: a gauge theory perspective},''
  \href{http://dx.doi.org/10.1142/S0218271816450024}{{\em Int. J. Mod. Phys.}
  {\bfseries D25} no.~13, (2016) 1645002},
\href{http://arxiv.org/abs/1604.05547}{{\ttfamily arXiv:1604.05547 [gr-qc]}}.

\bibitem{TRAUTMAN:1973aa}
A.~TRAUTMAN, ``Spin and torsion may avert gravitational singularities,'' {\em
  Nature Physical Science} {\bfseries 242} (03, 1973) 7 EP --.
  \url{http://dx.doi.org/10.1038/physci242007a0}.

\bibitem{PhysRevD.10.1066}
F.~W. Hehl, P.~von~der Heyde, and G.~D. Kerlick, ``General relativity with spin
  and torsion and its deviations from einstein's theory,''
  \href{http://dx.doi.org/10.1103/PhysRevD.10.1066}{{\em Phys. Rev. D}
  {\bfseries 10} (Aug, 1974) 1066--1069}.
  \url{https://link.aps.org/doi/10.1103/PhysRevD.10.1066}.

\bibitem{Shapiro:2001rz}
I.~L. Shapiro, ``{Physical aspects of the space-time torsion},''
  \href{http://dx.doi.org/10.1016/S0370-1573(01)00030-8}{{\em Phys. Rept.}
  {\bfseries 357} (2002) 113},
\href{http://arxiv.org/abs/hep-th/0103093}{{\ttfamily arXiv:hep-th/0103093
  [hep-th]}}.

\bibitem{Shie:2008ms}
K.-F. Shie, J.~M. Nester, and H.-J. Yo, ``{Torsion Cosmology and the
  Accelerating Universe},''
  \href{http://dx.doi.org/10.1103/PhysRevD.78.023522}{{\em Phys. Rev.}
  {\bfseries D78} (2008) 023522},
\href{http://arxiv.org/abs/0805.3834}{{\ttfamily arXiv:0805.3834 [gr-qc]}}.

\bibitem{Poplawski:2010kb}
N.~J. Pop?awski, ``{Cosmology with torsion: An alternative to cosmic
  inflation},'' \href{http://dx.doi.org/10.1016/j.physletb.2010.09.056,
  10.1016/j.physletb.2011.05.047}{{\em Phys. Lett.} {\bfseries B694} (2010)
  181--185}, \href{http://arxiv.org/abs/1007.0587}{{\ttfamily arXiv:1007.0587
  [astro-ph.CO]}}.
[Erratum: Phys. Lett.B701,672(2011)].

\bibitem{Banerjee:2018yyi}
R.~Banerjee, S.~Chakraborty, and P.~Mukherjee, ``{Late-time acceleration driven
  by shift-symmetric Galileon in presence of Torsion},''
\href{http://arxiv.org/abs/1802.04150}{{\ttfamily arXiv:1802.04150 [gr-qc]}}.

\bibitem{Parattu:2013gwa}
K.~Parattu, B.~R. Majhi, and T.~Padmanabhan, ``{Structure of the gravitational
  action and its relation with horizon thermodynamics and emergent gravity
  paradigm},'' \href{http://dx.doi.org/10.1103/PhysRevD.87.124011}{{\em Phys.
  Rev.} {\bfseries D87} no.~12, (2013) 124011},
\href{http://arxiv.org/abs/1303.1535}{{\ttfamily arXiv:1303.1535 [gr-qc]}}.

\bibitem{Davies:1974th}
P.~C.~W. Davies, ``{Scalar particle production in Schwarzschild and Rindler
  metrics},''
\href{http://dx.doi.org/10.1088/0305-4470/8/4/022}{{\em J. Phys.} {\bfseries
  A8} (1975) 609--616}.

\bibitem{Arcos:2005ec}
H.~I. Arcos and J.~G. Pereira, ``{Torsion gravity: A Reappraisal},''
  \href{http://dx.doi.org/10.1142/S0218271804006462}{{\em Int. J. Mod. Phys.}
  {\bfseries D13} (2004) 2193--2240},
\href{http://arxiv.org/abs/gr-qc/0501017}{{\ttfamily arXiv:gr-qc/0501017
  [gr-qc]}}.

\bibitem{Banerjee:2010yn}
K.~Banerjee, ``{Some Aspects of Holst and Nieh-Yan Terms in General Relativity
  with Torsion},'' \href{http://dx.doi.org/10.1088/0264-9381/27/13/135012}{{\em
  Class. Quant. Grav.} {\bfseries 27} (2010) 135012},
\href{http://arxiv.org/abs/1002.0669}{{\ttfamily arXiv:1002.0669 [gr-qc]}}.

\bibitem{Parattu:2015gga}
K.~Parattu, S.~Chakraborty, B.~R. Majhi, and T.~Padmanabhan, ``{A Boundary Term
  for the Gravitational Action with Null Boundaries},''
  \href{http://dx.doi.org/10.1007/s10714-016-2093-7}{{\em Gen. Rel. Grav.}
  {\bfseries 48} no.~7, (2016) 94},
\href{http://arxiv.org/abs/1501.01053}{{\ttfamily arXiv:1501.01053 [gr-qc]}}.

\bibitem{Chakraborty:2016yna}
S.~Chakraborty, ``{Boundary Terms of the Einstein?Hilbert Action},''
  \href{http://dx.doi.org/10.1007/978-3-319-51700-1_5}{{\em Fundam. Theor.
  Phys.} {\bfseries 187} (2017) 43--59},
\href{http://arxiv.org/abs/1607.05986}{{\ttfamily arXiv:1607.05986 [gr-qc]}}.

\bibitem{Arnowitt:1962hi}
R.~L. Arnowitt, S.~Deser, and C.~W. Misner, ``{The Dynamics of general
  relativity},'' \href{http://dx.doi.org/10.1007/s10714-008-0661-1}{{\em Gen.
  Rel. Grav.} {\bfseries 40} (2008) 1997--2027},
\href{http://arxiv.org/abs/gr-qc/0405109}{{\ttfamily arXiv:gr-qc/0405109
  [gr-qc]}}.

\end{thebibliography}\endgroup

\bibliographystyle{./utphys1}
\end{document}